   \documentclass[final,5p,times,twocolumn]{elsarticle}

\usepackage{amssymb}
\usepackage{dsfont}
\usepackage{amsmath}
\usepackage{float}
\usepackage{placeins,graphicx}

\DeclareMathOperator{\Tr}{Tr}
\newcommand{\mr}{\mathrm} 

\journal{Physics Letters B}

\begin{document}

\begin{frontmatter}



\title{On quantization of the SU(2) Skyrmions} 


\author[tfai]{D.~Jur\v ciukonis\corref{cor1}}
\ead{darius.jurciukonis@tfai.vu.lt}
\author[tfai]{E.~Norvai\v sas}
\ead{egidijus.norvaisas@tfai.vu.lt}
\cortext[cor1]{Corresponding author}

\address[tfai]{Vilnius University Institute of Theoretical Physics and Astronomy, Go\v stauto 12, Vilnius 01108, Lithuania}

\begin{abstract}
There are two known approaches for quantizing the SU(2) Skyrme model, the semiclassical and canonical quantization. The semiclassical approach does not take into account the non-commutativity of velocity of quantum coordinates and the stability of the semiclassical soliton is conveniently ensured by the symmetry breaking term. The canonical quantum approach leads to quantum mass correction that is not obtained in the semiclassical approach. In this letter we argue that these two approaches are not equivalent and lead to different results. We show that the resulting profile functions have the same asymptotic behaviour, however their shape in the region close to the origin is different.
\end{abstract}


\end{frontmatter}


\section{Introduction}
\label{Sect_intr}

The Skyrme topological soliton model is a nonlinear field theory, with localized finite energy soliton solutions \cite{Skyrme61,Skyrme62}. The first comprehensive phenomenological application of the model to baryons was the semiclassical calculation of the static properties of the nucleon \cite{Adkins}. The low-energy QCD in the large color limit \cite{Witten} is argued to describe baryons as solitons in the weakly coupled phase of mesons which was the original idea of \cite{Skyrme61} and \cite{Adkins}. The original model was defined for a unitary field $U(\mathbf{x},t)$ that belongs to the fundamental representation of SU(2). The boundary constraint $U\rightarrow \mathds{1}$ as $\left\vert \mathbf{x}\right\vert \rightarrow\infty $ implies that the unitary field represents a mapping from $S^{3}\rightarrow S^{3}$. The integer valued winding number which classifies the solitonic sectors of the model was interpreted to be the baryon number. The semiclassical quantization of model has proven to be useful for baryon phenomenology. However the semiclassically-treated SU(2) model was shown to have an instability in calculating the energy functional \cite{Bander, Braaten}. The stability and correct asymptotic behaviour of solutions can be achieved by introducing an additional symmetry breaking term. The alternative stabilization of the quantum SU(2) Skyrme model has been obtained by quantizing the soliton quantum canonically in collective coordinate approach \cite{Fujii87,Acus98}. The non-commutativity of canonical momenta in a Hamiltonian system leads to non-commutativity of velocities of the canonical coordinates (collective coordinates) which can not be ignored. It was shown that the procedure of the canonical quantization contributes to the appearance of new terms in the explicit form of the Lagrangian of the model. These terms are interpreted as quantum corrections to the mass of the soliton (`quantum mass corrections') that restore the stability of the solitons that is lost in the semiclassical approach \cite{Acus98}. The purpose of the present letter is to show that the quantum mass corrections of the soliton are important in ensuring the stability of the quantum solitons and realize Skyrme's original conjecture that `the mass (of the meson $m_{\pi}$) may arise as a self-consistent quantal effect. This point will not be followed here, but when, for calculation purposes, we want to allow phenomenologically for a finite mass this will be done by adding to $L$ a term (proportional to $m^2_{\pi}$)' \cite{Skyrme62}. We find stable quantum solitons by varying the complete quantum energy functional for nucleon. The stability is ensured by the consequence of iterative calculations. The shapes of quantum solitons with different stabilising terms are demonstrated in Fig.~\ref{figure2}. We do not consider quantum soliton with quantum numbers of $\Delta$ resonances because there are no known stable solutions for the quantum SU(2) Skyrme model defined in fundamental representation. The stable quantum solutions for $\Delta$ exist in the generalized SU(2) Skyrme model which is defined for higher representations \cite{Acus98} or in the SU(3) Skyrme model \cite{Jurc13}.  After some preliminary definitions in Section~\ref{Sect_clas} below, the main part of this paper is organized as follows. In Section~\ref{Sect_quant} the quantum Skyrme model is constructed \textit{ab initio} in the collective coordinates framework and canonicaly quantized. The structure of energy functional is derived. Section~\ref{Sect_num} contains numerical results and summarizing discussion.

\section{Classical skyrmion}
\label{Sect_clas}

The SU(2) Skyrme model is conveniently defined via the chirally symmetric Lagrangian density
\begin{equation}
\mathcal{L}_{\mr{Sk}}=-\frac{f_{\pi }^{2}}{4}\Tr\{\mathbf{R}_{\mu }%
\mathbf{R}^{\mu }\}+\frac{1}{32e^{2}}\Tr\{[\mathbf{R}_{\mu },%
\mathbf{R}_{\nu }][\mathbf{R}^{\mu },\mathbf{R}^{\nu }]\} \,,  \label{G1}
\end{equation}%
which is written in terms of the su(2)-valued right chiral current $\mathbf{R}_{\mu }=\left(
\partial _{\mu }U\right) U^{\dagger }$. Here $f_\pi$ and $e$ are model parameters, whose values are constrained by fitting with the experimental data. 
The chiral symmetry breaking term of Lagrangian density is defined by
\begin{equation}
\mathcal{L}_{\mr{SB}}=-{\mathcal{M}}_{\mr{SB}}=-\text{\thinspace }\frac{f_{\pi}^{2}}{4} m_{0}^{2}\,\textrm{Tr}\left\{ U+U^{\dagger }-2\cdot \mathds{1}\right\},  \label{G2}
\end{equation}
where $m_0$ is the third model parameter. 

The classical static soliton (Skyrmion) is obtained by employing the spherically symmetric hedgehog ansatz
\begin{equation}
U_{0}\left( \hat{x},F(r)\right) =\exp i(\mathbf{\sigma }\cdot \hat{x})F(r) ,  \label{G3}
\end{equation}
where $\mathbf{\sigma }$ are Pauli matrices and $\hat{x}$ is the unit vector. With this ansatz the classical Lagrangian density reduces to the following simple form
\begin{align}
\mathcal{L}_{\mr{cl}}(F(r)) =& -\mathcal{M}_{\mr{Sk}}(F(r))-\mathcal{M}_{\mr{SB}}(F(r)) \notag  \\ 
 =& -{\frac{f_{\pi }^{2}}{2}}\Bigl(F^{\prime 2}+{\frac{2}{r^{2}}}\sin ^{2}\!F\Bigr)  
 -{\frac{1}{2e^{2}}}{\frac{\sin ^{2}\!F}{r^{2}}}\Bigl(2F^{\prime 2}
 +{\frac{\sin ^{2}\!F}{r^{2}}}\Bigr) \notag
\\ 
 & -f_{\pi }^{2}m_{0}^{2}(1-\cos\!F)\,,  \label{G4}
\end{align}
which defines the energy of the Skyrmion, ${E}_{0} = -4\pi\int\mathcal{L}_{\mr{cl}}(F(r))\,r^2\mathrm{d} r$.
%
%
Variation of \eqref{G4} leads to a differential equation for the profile function $F(r)$. The standard boundary conditions for a Skyrmion are $F(0)=\pi$, $F(\infty )=0$.

\section{Quantization of the Skyrmion}
\label{Sect_quant}

The standard (semiclassical) approach to quantize the rotational zero modes of the Skyrmion yielding multiplets with equal spin and isospin in each multiplet was presented in \cite{Adkins,Witten}. This approach treats the soliton as a rigid body and does not take into account the non-commutativity of quantum coordinates and velocities. The canonical quantization treats the quantum variables canonically and leads to new quantum terms in the explicit of the Lagrangian that are interpreted as dynamically generated quantum mass correction. This approach was presented in \cite{Fujii87} and further developed in \cite{Acus98}. The details are as follows.

The quantum field $U$ can be written in a form with temporal and spatial parts separated explicitly 
\begin{equation}
U(\hat{x},F(r),\mathbf{q}(t))=A(\mathbf{q}(t))U_{0}\left( \hat{x}
,F(r)\right) A^{\dagger}(\mathbf{q}(t)) \,,  \label{B1}
\end{equation}
where $U_0$ is the classical field and $A(\mathbf{q}(t))$ is a matrix specified by three real independent parameters, generalized quantum coordinates $q^{k}(t)$.
The Lagrangian (\ref{G1}) is considered quantum mechanically \textit{ab
initio}. Thus the canonical commutation relation $\left[
p_{k},q^{l}\right] =-i\delta _{kl}$ \ for generalized coordinates $q^{l}$
and conjugate momenta $p_{k}$ is required to hold. In such a way the generalized coordinates $q^{k}(t)$ and velocities are ought to satisfy the commutation relations
\begin{equation}
\left[ \dot{q}^{k},q^{l}\right] =-if^{kl}(q)\,,  \label{B2}
\end{equation}%
where the form of the tensor $f^{kl}(q)$  will be determined below. The temporal derivatives are calculated by employing the usual Weyl ordering, and the operator ordering is fixed by the form of the Lagrangian (\ref{G1}) 
without further ordering ambiguity. For the derivation of the canonical momenta it is sufficient to restrict the consideration to the terms of second order in velocities (the terms of first order vanish). This leads to
\begin{equation}
L_{\mr{Sk}}\approx \,\frac{1}{2}\,\dot{q}^{\alpha }g_{\alpha \beta }(q,F)\dot{q}^{\beta }+\left[ (\dot{q})^{0}\mathrm{-order\,\,term}\right] , \label{B5}
\end{equation}%
where the metric tensor takes the form 
\begin{equation}
g_{\alpha \beta }(q,F)=-C_{\alpha }^{\prime(M)}(q)(-1)^{M}a(F)\delta _{M,-M^{\prime }}C_{\beta }^{\prime (M^{\prime })}(q)\,.  \label{B6}
\end{equation}
Here $C_{\alpha }^{\prime (M)}(q)$ are functions of quantum coordinates $q$. Their explicit form depends on the chosen parametrization of the SU(2) group. However the explicit form does not appear anywhere in the calculations. For details on these functions we refer to \cite{Acus98}.

The canonical commutation relations $\left[ p_{\beta },q^{\alpha }\right] =-i\delta _{\alpha \beta }$ then yield the explicit expression for the functions $f^{\alpha \beta }(q)=g_{\alpha \beta }^{-1}(q,F)$. Next, by substituting (\ref{B1}) into the Lagrangian density (\ref{G1})  and after some lengthy manipulation and integration over the space variables the complete expression of the quantum Skyrme model Lagrangian are obtained
\begin{equation}
L=-M_{\mr{cl}}-\Delta M +\frac{1}{2a(F)}\hat{J}^{\prime 2}\,,  \label{C13}
\end{equation}
where $\Delta M$ is the (negative) quantum mass correction
\begin{equation}
\Delta M =-\frac{2\pi }{a^{2}(F)}\int r^{2}\mathrm{d}r\sin ^{2}F%
\left[ f_{\pi }^{2}+\frac{1}{2e^{2}}\left( 2F^{\prime 2}+\frac{\sin ^{2}F}{%
r^{2}}\right) \right], \label{C14}
\end{equation}
and $\hat{J}_{(M)}^{\prime }$ are the angular momentum operators
\begin{equation}
\hat{J}_{(M)}^{\prime }=\frac{i}{2}\left\{ p_{\alpha },C_{(M)}^{\prime \alpha }(q)\right\} \label{C6}
\end{equation}%
satisfying the standard SU(2) commutation rules, and $C_{(M)}^{{\prime }\alpha}(q)$ is the reciprocal matrix to $C_{\alpha }^{\prime (M)}(q)$.  
The generalized method of quantization on a curved space developed by Sugano {\it et all}. \cite{Sugano} allows to write the energy functional of the quantum Skyrmion for a state with fixed spin and isospin $\ell $ in this form
\begin{equation}
E(\ell ,F)=M_{\mathrm{cl}}(F)+\Delta M(F)+\frac{\ell (\ell +1)}{2a(F)} \,, \label{C17} 
\end{equation}
where $a(F)$ is the quantum momentum of inertia of the Skyrmion
\begin{equation}
a(F) = \frac{1}{e^{3}f_{\pi }}\tilde{a}(F)=\frac{1}{e^{3}f_{\pi }}\frac{8\pi }{3}\int \mathrm{d}\tilde{r}\tilde{r}^{2}\sin ^{2}F \,\Big(1+F^{\prime 2}+\frac{1}{\tilde{r}^{2}}\sin ^{2}F\Big) \,.  \label{B8}
\end{equation}
Notice that it differs from the mechanical moment of inertia of the classical Skyrmion.
%
The expression \eqref{C17} is the quantum version of the mass formula of the
Skyrme model, which differs from the semiclassical one by the appearance of the additional (negative) quantum correction $\Delta M(F)$. 
The variation of the energy functional $\frac{\delta E(F)}{\delta F}=0$ of the quantum Skyrmion for states with given $\ell $ leads to an integrodifferential equation for the profile function $F(r)$ with the same boundary conditions as in the classical case, $F(0)=\pi$ , $F(\infty )=0$. At large distances 
the asymptotic solution takes the form
\begin{equation}
F(\tilde{r})=k\left( \frac{\tilde{m}^{2}}{\tilde{r}}+\frac{1}{\tilde{r}^{2}}%
\right) \exp (-\tilde{m}\tilde{r}) \,,  \label{F7}
\end{equation}%
where the quantity $\tilde{m}^{2}$ is defined by
\begin{equation}
\tilde{m}^{2}=-\frac{e^{4}}{3\tilde{a}(F)}\left\{\frac{2\ell (\ell +1)+3}
{\tilde{a}(F)}+8\Delta \tilde{M}(F)\right\}+\tilde{m}_{0}^{2} \,.  \label{f11}
\end{equation}
The expressions above are given in terms of the dimensionless parameters $\tilde{r}=e f_{\pi }r$ and $\tilde m = m / e f_\pi$. The Eq.~\eqref{F7} and \eqref{f11} are in the agreement with Eq.~(16) of \cite{Braaten}. The additional negative $\Delta M$ quantum mass correction appear from noncommuting of quantum coordinates and quantum velocities.

\section{Numerical results and discussion} 
\label{Sect_num}

The Skyrmion is quantized within the Bohr framework by requiring the spin to be half-integer valued and taking values $\ell=1/2$ for the nucleon $(N)$, and $\ell = 3/2$ for the delta $(\Delta)$. 
 
In the rigid body approach the spining Skyrmion is treated as a rigid quantum rotator, namely it is assumed that the soliton does not deform when it spins. In such a way $N$ and $\Delta$ are both described by the same classical profile function obtained by minimizing the classical energy functional \eqref{G4}. 
Such treatment allows to calculate energies of the spinning Skyrmion with an arbitrary tower of spin-isospin. However such treatment of the model is not entirely physical as it suffers from some ambiguous artefacts, as it was pointed out in \cite{Bander,Braaten,RSWW}.

The classical static soliton mass has three terms $M_2$, $M_4$ and $M_{\text{SB}}$ with different dependence on the scaling parameter. These terms correspond to the 2nd, 4th order and chiral symmetry breaking terms in the Lagrangian. Due to Derrick theorem the stable solitonic solutions can exist if $M_4 = M_2 + M_{\text{SB}}$. In the time-dependent case the canonical quantization gives rise to the kinetic term $\ell(\ell + 1)/2a(F)$ and quantum mass correction $\Delta M(F)$, which appear in total energy functional and have quantum momentum of inertia $a(F)$ in the denominator. From \eqref{B8} we can see that $a(F)$ has two parts which originate from the 2nd and 4th order terms. These parts have different dependence on the scaling parameter. The total energy functional can then be formally expressed as infinite series of the scaling parameter and relations of Derrick type can not be obtained.  

\begin{figure}[t]
\centering
\includegraphics[width=0.47\textwidth]{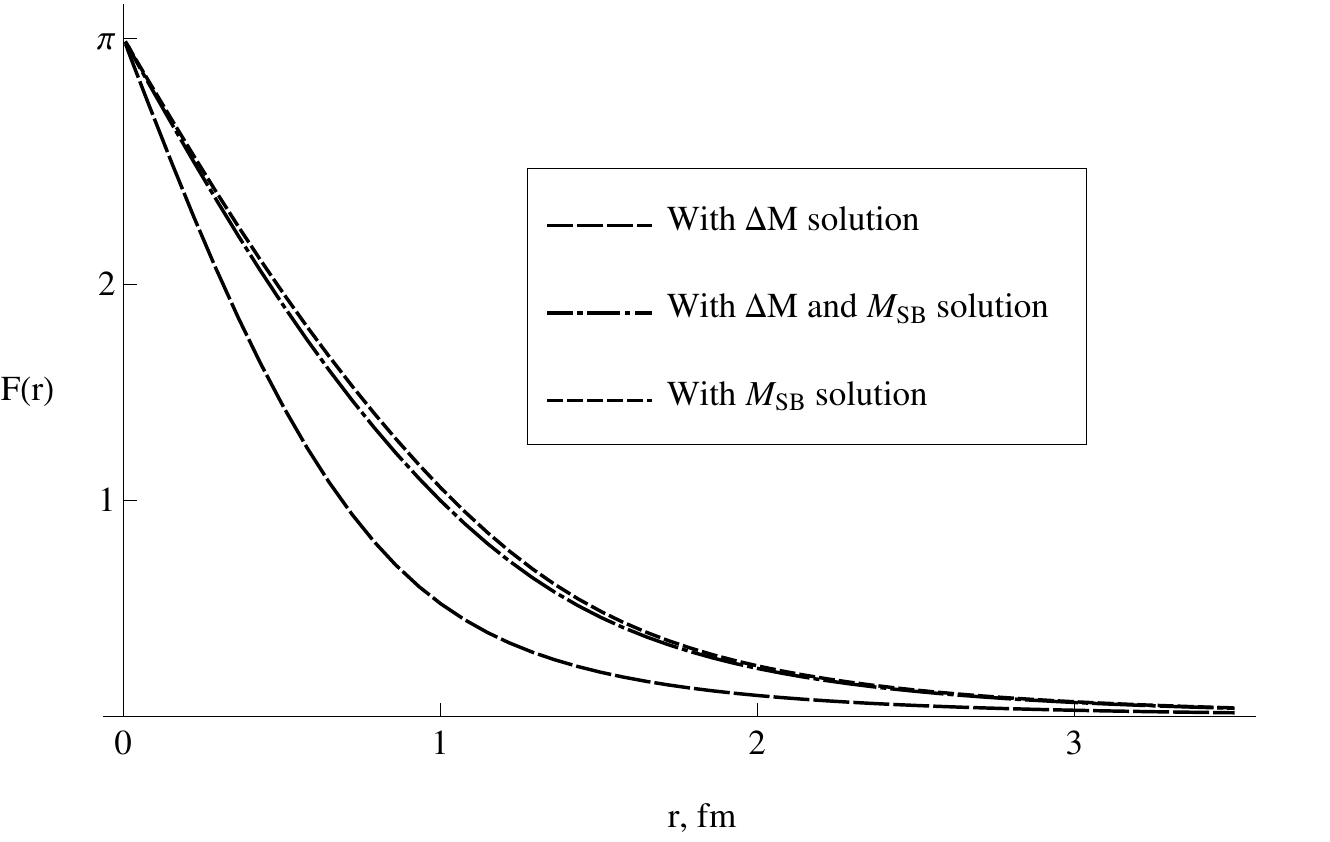}
\caption{Nucleon profile function, $F(r),\; r$[fm].}
\label{figure1}
\end{figure}

A more realistic approach is to allow the spinning Skyrmion to deform within the spherically symmetric hedgehog ansatz. In such a way it is no longer a {\it rigid} quantum rotator, rather a {\it soft} quantum rotator what is in agreement with the canonically quantized Lagrangian \eqref{G1}. In this approach the soliton profile function $F(r)$ is determined by minimization of the energy functional \eqref{C17}, which is an integro-differential equation and needs to be solved by means of the iterative calculations. In this approach a localized finite energy solution (quantum soliton) can only exist if $\tilde{m}^2 > 0$, as can be easily deduced from \eqref{F7} and \eqref{f11}. This constraint puts an upper bound on the isospin $\ell$ and also severely constrains the allowed range of values for the Skyrme parameter $e$. Furthermore, the experimental value of the pion mass, the standard values of parameters $f_\pi$ and $e$ in this approach do not lead physically meaningful results, especially for the $\Delta$, for which no spinning solution exists within these values of the parameters. This is particularly important in the semiclassical approach when the $\Delta M$ term is absent. Hence the standard Skyrme parameters are considered to be an artefact of the rigid body approximation. 

\begin{table*}[t]
\begin{center}
\begin{tabular}{|c|c|c|c|c|}
\hline \hline
Parameters & \,With $\Delta M$ only\, & \,With $M_{\mathrm{SB}}$ only\, &  With $\Delta M$ and $M_{\mathrm{SB}}$ & \,Exp.\,  \\
\hline
$e$ & 4.96 & 4.43 & 4.08 & ---  \\
\hline
$f_{\pi}$ [MeV] & 68.3 & 48.3 & 50.9 & 92.2   \\
\hline
$m_0$ [MeV] & --- & 170.7 & 120.7 & ---  \\
\hline
$\langle r^2\rangle ^{1/2}$ [fm] & 0.54  & \small{Input} & \small{Input} & 0.81  \\
\hline
$m_N$ [MeV] & \small{Input} & \small{Input} & \small{Input} & 939  \\
\hline
$m$ [MeV]& \small{Input} & \small{Input} & \small{Input} & 137.7  \\
\hline
$\mu_p $ & 1.64 & 2.69 & 2.97 & 2.79  \\
\hline
$\mu_n $ & -1.06 & -1.93 & -2.32 & -1.91  \\
\hline
$g_A $ & 0.89 & 1.06 & 1.24 & 1.26  \\
\hline\hline
\end{tabular}
\end{center}
\caption{Model parameters for nucleons as stable solitonic solutions. Calculated magnetic moments for proton $\mu_p$ and neutron $\mu_n$, and axial coupling constant $g_A$.} \label{table1}
\end{table*}

Physically meaningfull results that agree reasonably well with experimental data can be obtained by allowing $f_\pi$, $e$ and $m_0$ to be arbitrary parameters of the models that are determined by the fitting the model to some selected physical properties of $N$ and $\Delta$ \cite{Acus98,Acus01,Battye}. A higher precision of the model may further be achieved by taking into account higher order quantum effects or by employing some specific approaches, for example by deforming the hedgehog ansatz \cite{Battye,WWS,Dorey}, or by considering multi-Skyrmions \cite{Battye1,multi}. Interestingly, all of these approaches find the parameter $m_0$ of the symmetry breaking term \eqref{G2}, which is conveniently identified with the pion mass, to be significantly bigger than the experimentally observed value. This observation is in a perfect agreement with the appearance of the quantum mass correction $\Delta M$ in the canonically quantized approach. This mass correction, which has a negative absolute value, contributes to the effective mass of the pion $\tilde{m}$ \eqref{f11} thus lowering the value of $m_0$ close to the experimental data. Let us next give the details of our numerical calculations.

\begin{figure}[t]
\centering
\includegraphics[width=0.47\textwidth]{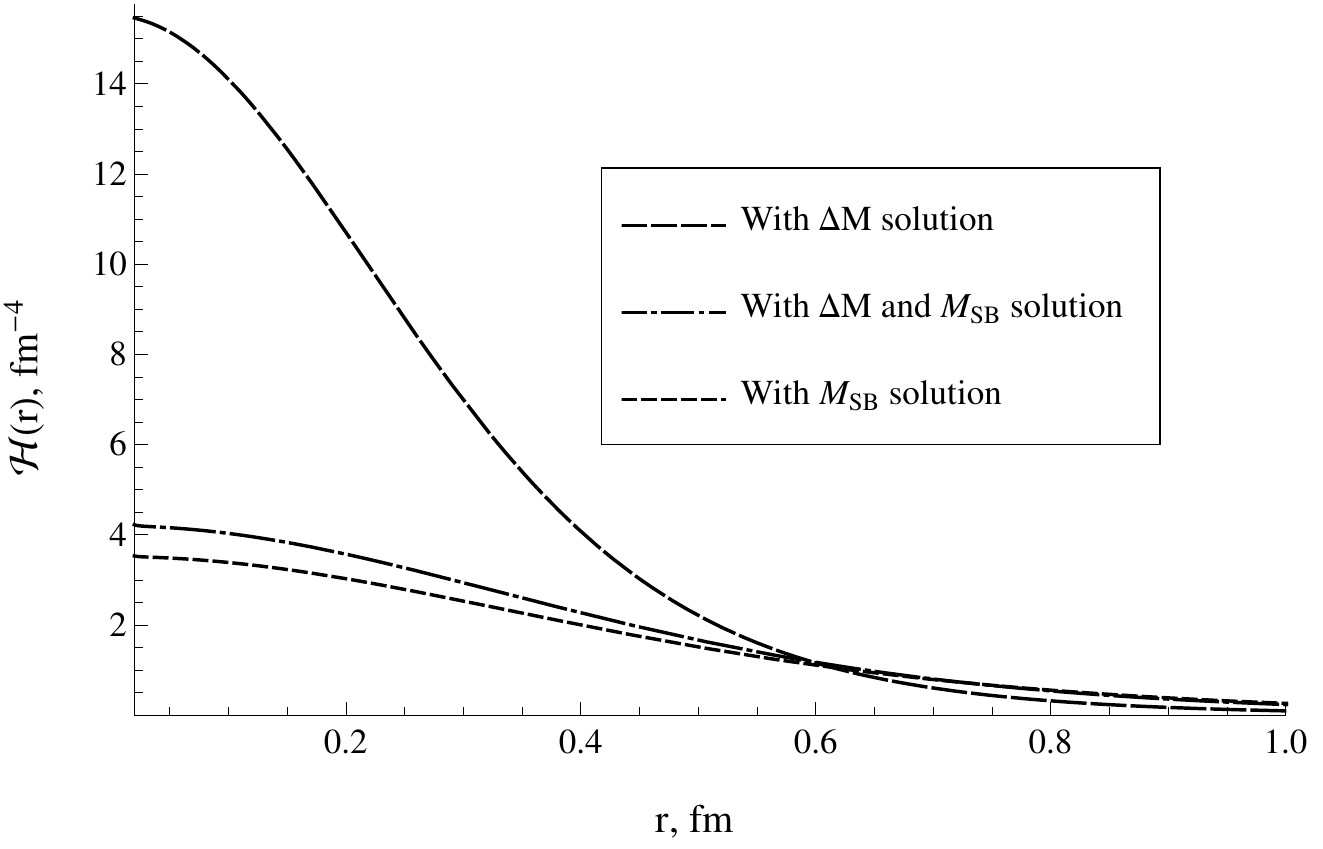}
\caption{Nucleon mass distribution, $\mathcal{H}(r)$[fm$^{-4}$], $r$[fm].}
\label{figure2}
\end{figure}
We have considered three different approaches, the canonical quantization without the symmetry breaking term, the semiclassical quantization with the symmetry breaking term, and the canonical quantization with the symmetry breaking term. These approaches correspond to second, third and fourth columns of Table~\ref{table1}, respectively. 

For each case we have found a profile function $F(r)$ minimizing energy functional \eqref{C17} with appropriate terms included or excluded ($\Delta M$ and $M_{\rm SB}$). The model parameters were determined by fixing the nucleon mass to $E(F)=m_{N}=939$ MeV, the asymptotic behaviour
of the nucleon mass distribution (the asymptotic nucleon mass) to $m=m_{\pi }=137.7$ MeV, and requiring the nucleon isoscalar mean square radius to be $\langle r^{2}\rangle=0.657$ fm (for the cases with $M_{\rm SB}$ term included). The isoscalar mean square radius is given by
\begin{equation}
\bigl\langle r^{2}\bigr\rangle\!=\!-\frac{2}{\pi e^{2}f_{\pi }^{2}}\int
\!\tilde{r}^{2}F^{\prime }\sin ^{2}\!F\mathrm{d}\tilde{r} \,.  \label{radius}
\end{equation}

Numerical calculations of the integro-differential
equation (\ref{C17}) are performed in the following way.
\begin{enumerate}
\item Using the classical profile function and set of empirical baryon observables (the
nucleon mass (\ref{C17}), the asymptotic nucleon mass (\ref{f11}) and the isoscalar radius (\ref{radius})) we fit three model
parameters $f_\pi$, $e$ and $m_0$ and calculate all required integrals in
the quantum equation (\ref{C17}).

\item Using the known asymptotic solution (\ref{F7}) (and its
derivative) we adopt a simple procedure solving the
differential equation and find the first approximation of the
quantum solution $F^{(1)}(\tilde r)$ and the constant $k^{(1)}$ in
(\ref{F7}). \label{item}

\item  The obtained function $F^{(1)}(\tilde r)$ is used to
recalculate $f_\pi, e, m_0$ and the integrals. The procedure described
in item~\ref{item} can be used again to get the second
approximation to the quantum solution $F^{(2)}(\tilde r)$ and the
constant $k^{(2)}$.

\item This procedure is iterated until the convergent solution
and the parameters $f_\pi, e, m_0$ as well as stable values of
$M_{\text{cl}}, \Delta M, a,\tilde{m}$ are
obtained. The self-consistent set then can be used to calculate
numerous phenomenologically interesting quantities.
\end{enumerate}

The explicit expressions for calculating magnetic moments $\mu_p $, $\mu_n $   and the axial coupling constant $g_A$ were taken from \cite{Acus98}.
The nucleon profile function for each case is shown in Fig.~\ref{figure1}, the corresponding radial mass distributions of the nucleon are shown in Fig.~\ref{figure2} (see \cite{Acus01}). The overall contribution of the $\Delta M$ term to the energy of the state is negative and can be interpreted as appearing due to self-interactions of the quantum pion cloud and may be understood as a dynamically generated mass term. In contrast, the overall contribution of the $M_{\rm SB}$ term is strictly positive. In such a way the profile function for the case with the $\Delta M$ term only is much more localised than in the case with the $M_{\rm SB}$ term only. Hence these two terms have different physical origin and can not understood equivalent to each other in any way. Interestingly, inclusion of both $\Delta M$ and $M_{\rm SB}$ terms leads to a profile function which is very similar to the one of the case with the $M_{\rm SB}$ term only. Although being of a similar form, they are solutions of significantly different equations, which translates into different model parameters as can be seen in Table~\ref{table1}. Here we see that the approach with both terms included leads a value of $m_0$ close to $m_\pi$ as we have mentioned earlier. Furthermore, these approaches lead to significantly different magnetic moments for proton $\mu_p$ and neutron $\mu_n$, and axial coupling constant $g_A$, see Table~\ref{table1}. Hence these approaches would lead to different shape of the nucleon form factor, and thus can not be treated at an equal footing. The canonical quantization ensures existence of self-consistent quantum Skyrmions and there are no mathematical or physical reasons to remove quantum mass correction from the Skyrme model Lagrangian.

\section*{Acknowledgements} 

The authors thank Paul Sutcliffe and Vidas Regelskis for valuable discussions and suggestions.

\end{document}